\documentclass[superscriptaddress,twocolumn]{revtex4}
\usepackage{graphicx}
\usepackage{color}

\renewcommand{\vec}[1]{\mbox{\boldmath $#1$}}

\begin{document}

\title{
Subbarrier fusion reactions of an aligned deformed nucleus} 

\author{K. Hagino}
\affiliation{ 
Department of Physics, Kyoto University, Kyoto 606-8502,  Japan} 

\author{S. Sakaguchi}
\affiliation{ 
Department of Physics, Kyushu University, Fukuoka 819-0395,  Japan} 


\begin{abstract}
We discuss heavy-ion fusion reactions of a well-deformed odd-mass 
nucleus at energies around the Coulomb barrier. To this end, we 
consider the $^{16}$O+$^{165}$Ho reaction and take into account the 
effect of deformation of $^{165}$Ho using the orientation average formula. 
We show that fusion cross sections are sensitive to magnetic substates 
of the target nucleus, 
and cross sections for the side collision, which are relevant to 
a synthesis of superheavy elements, may be enhanced by a factor of around 
two by aligning the deformed target nuclei. We also discuss the effect of 
alignment on the fusion barrier distribution. 
We find that, for a particular choice of alignment, the shape of 
barrier distribution becomes similar to a typical shape of 
barrier distribution 
for a deformed nucleus with a negative hexadecapole 
deformation parameter, $\beta_4$, 
even if the intrinsic $\beta_4$ is positive in the target nucleus. 
\end{abstract}


\maketitle

\section{Introduction}

It has been well known that nuclear deformation of the colliding nuclei 
plays an important role in heavy-ion fusion reactions 
\cite{BT98,DHRS98,HT12,Back14,MS17}. 
For medium-heavy systems, such 
as $^{16}$O+$^{154}$Sm, fusion cross sections are largely enhanced 
at energies around the Coulomb barrier 
compared to a prediction of a simple 
potential model. This has been interpreted as a result of 
a distribution of the Coulomb barriers, that depend 
upon the orientation angle 
of the deformed target nucleus. Since fusion cross sections have an exponential 
dependence on the incident energy at energies below the Coulomb barrier, 
fusion cross sections 
can be enhanced by orders of magnitude due to the contribution of 
the configurations for which the Coulomb barrier is lowered than the 
original barrier. 
The barrier distribution has been investigated 
experimentally for several systems \cite{DHRS98,RSS91,LDH95}, 
and this picture has been well established by now. 

The nuclear deformation plays an important role also in fusion reactions 
in massive systems, that is, those used to synthesize superheavy nuclei. 
For prolately deformed nuclei, a compact configuration is realized at the 
touching point when a projectile nucleus approaches from the 
direction of the shorter axis of the target nucleus, that is, the side 
collision. This makes it a big advantage to overcome an inner barrier and 
form a compound 
nucleus \cite{Hinde18,Hinde95,Hinde96,Nishio08,Nishio00,RGT07,Hagino18}. 
The notion of compactness has recently been confirmed experimentally 
by comparing the measured barrier distribution and the excitation 
function of evaporation residue cross sections 
for the $^{48}$Ca+$^{248}$Cm 
system \cite{Tanaka18}. 
Moreover, employing the concept of the compactness, 
the so called hugging fusion was proposed for fusion between 
deformed nuclei with negative hexadecapole deformation, such as 
$^{150}$Nd+$^{150}$Nd, for which the touching configuration is compact when 
the symmetry axis of each nucleus is perpendicular 
to each other \cite{Iwamoto96}. (Notice, however, that 
the effective inner barrier will be considerably high for such 
symmetric systems \cite{Cap14} 
and it would be extremely difficult to synthesize superheavy 
elements with hugging fusion.) 

In many experiments for fusion of a deformed nucleus, an even-even nucleus 
has been chosen as a target nucleus, partly because 
the ground state has zero spin and thus the theoretical 
treatment is easy. However, a finite spin of odd-mass nuclei may 
bring an interesting insight into dynamics of heavy-ion fusion 
reactions \cite{Colucci19} (see also Ref. \cite{Piasecki19}). 
As a matter of fact, fusion of an aligned/polarized nucleus has been 
theoretically 
investigated in the 1980s and 1990s and it has been demonstrated 
that fusion cross sections are significantly altered by aligning the odd-mass 
nuclei \cite{SG81,JS85,FG86,MGCA92,Christley94}. 
Notice that this effect would be important in fusion in astrophysical 
environments under 
a strong magnetic field, which leads to a natural polarization of 
colliding nuclei. 
Experimentally, fusion of an aligned light nucleus, $^{23}$Na, has been 
measured \cite{Butsch87,Reich94,FGT92}. For heavier deformed nuclei, a 
measurement of fusion cross sections for the $^{16}$O+$^{165}$Ho system was 
planned \cite{Christley94}, even though 
the actual experiment has not yet been
performed so far \cite{Walker}. 

In this paper, we revisit the problem of fusion of an aligned deformed 
nucleus. In the previous studies, the effect of alignment was discussed 
only in terms of fusion cross sections as well as tensor analyzing powers, 
and the effect on fusion barrier distributions has yet to be
investigated. Notice that the shape of fusion barrier distribution is 
sensitive to the details of nuclear deformation, and 
one may gain a deeper insight into the reaction dynamics of a deformed 
nucleus by analyzing the fusion barrier distributions. 
Moreover, the effect of alignment has never been discussed in connection 
to fusion for superheavy elements, which may be important for future 
experiments to synthesize new superheavy elements. 
Given the new experiment for the barrier distribution for systems 
relevant to superheavy elements \cite{Tanaka18}, we consider that 
it is timely to revisit this problem now. 

The paper is organized as follows. In Sec. II, we summarize the theoretical 
framework to describe fusion cross sections for an aligned deformed target. 
To this end, we consider fusion of a well-deformed nucleus, for which 
fusion cross sections are approximately given as an average of 
fusion cross sections for fixed orientation angles. 
In Sec. III, we apply the formula to the $^{16}$O+$^{165}$Ho reaction, and 
discuss the effect of alignment on the fusion cross sections and 
the fusion barrier distribution. We also 
discuss an implication of the effect of alignment 
for fusion for superheavy nuclei. 
We then summarize the paper in Sec. IV. 

\section{Fusion cross sections for an aligned deformed target} 

We consider fusion between an inert projectile and a well-deformed 
target nucleus. 
In this case, the relative motion between the colliding nuclei couples 
to the rotational motion of the deformed target. 
We take a rigid rotor model to describe the wave 
functions for the ground state rotational band of the target nucleus. 
That is, 
for the state with the angular momentum $I$ and its $z$-component $M$, 
the wave function reads \cite{Rowe70}. 
\begin{eqnarray}
\Psi_{IKM}(\Omega)&=&\sqrt{\frac{2I+1}{16\pi^2(1+\delta_{K,0})}}\,
\left(D^I_{MK}(\Omega)\phi_K(\xi)\right. \nonumber \\
&&\left. +(-1)^{I+K}D^I_{M-K}(\Omega)\phi_{\bar{K}}(\xi)\right),
\end{eqnarray}
where $K$ is the $K$-quantum number, that is, the projection of the 
angular momentum on the body-fixed frame. 
Here, we have assumed that the deformed nucleus 
has axially symmetric shape so that the $K$-quantum number is conserved. 
$D^I_{MK}(\Omega)$ is the Wigner D-function, 
in which $\Omega=(\phi,\theta,\chi)$ denotes the Euler angles. 
$\phi_K(\xi)$ is the intrinsic wave function, where $\xi$ is the 
intrinsic coordinate, and $\phi_{\bar{K}}$ is the time-reverse 
of $\phi_{K}$. 

To simplify the angular momentum coupling, we employ the 
iso-centrifugal approximation \cite{HT12}. In this approximation, 
one transforms the whole system to the rotating frame where the $z$ 
axis is along the direction of the relative motion, $\vec{r}$, 
at every instant. 
The interaction between the projectile and the target 
nuclei in this approximation then reads \cite{HT12,HRK99}, 
\begin{equation}
V(r,\theta)=V_N(r,\theta)+V_C(r,\theta), 
\label{pot}
\end{equation}
with the nuclear potential given by, 
\begin{equation}
V_N(r,\theta)=-\frac{V_0}{1+\exp\left[\left(r-R_0-R_T\sum_\lambda\beta_\lambda
Y_{\lambda 0}(\theta)
\right)/a\right]},
\end{equation}
and the Coulomb potential given by, 
\begin{eqnarray}
V_C(r,\theta)&=&\frac{Z_PZ_Te^2}{r}\left(1
+\frac{3}{5}\,\frac{R_T^2}{r^2}
\left(\beta_2+\frac{2}{7}\sqrt{\frac{5}{\pi}}\beta^2\right)Y_{20}(\theta)\right. \nonumber \\
&&\left.
+\frac{3}{9}\,\frac{R_T^4}{r^4}
\left(\beta_4+\frac{9}{7}\beta^2\right)Y_{40}(\theta)\right).
\end{eqnarray}
Here, $\beta_\lambda$ are the deformation parameters of the deformed 
target and $\theta$ denotes the angle between the symmetry axis of the 
deformed target and the relative coordinate, $\vec{r}$. 
For a prolately deformed target, $\theta=0$ ($\theta=\pi/2$) 
corresponds to 
the case where the projectile nucleus approaches from the longer (shorter) 
axis of 
the target and refers to as 
the tip (side) collision.  
$R_T$ and $Z_T$ are the radius and the charge number of the target nucleus, 
respectively, and $Z_P$ is the charge number of the projectile nucleus. 
Here, we have assumed a Woods-Saxon shape for the nuclear potential, and 
expand the Coulomb potential up to the second order of $\beta_2$ and the 
first order of $\beta_4$. 
Notice that in the iso-centrifugal approximation 
the value of $M$ is conserved during the reaction, 
since the coupling potential, Eq. (\ref{pot}), does not change 
the $z$-component of the angular momentum  \cite{HT12,HTBB95}. 
The fusion cross sections can thus be labeled with $M$. 

In addition to the iso-centrifugal approximation, 
we take the sudden tunneling approximation by setting 
the rotational energy of the target nucleus to be zero. In this 
approximation, together with the iso-centrifugal approximation, 
fusion cross sections are given as a weighted average of 
cross sections for fixed values of the angle $\theta$, with the weight 
factors given by the square of the ground state wave function of the 
target nucleus \cite{HT12,SG81,TAB92,NBT86,Wong73}. 
That is, fusion cross sections for a magnetic substate $M$ read, 
\begin{eqnarray}
\sigma_{\rm fus}^{(M)}(E)
&=&\int d\Omega \,|\Psi_{I_{0}K_{0}M}(\Omega)|^2\,
\sigma_{\rm fus}(\theta), \\
&=&\int^{2\pi}_0d\phi\int^\pi_0\sin\theta d\theta \int^{2\pi}_0d\chi \nonumber \\
&&\times|\Psi_{I_{0}K_{0}M}(\Omega)|^2\,
\sigma_{\rm fus}(\theta), 
\label{cross0}
\end{eqnarray}
where $I_0$ and $K_0$ are the value of $I$ and $K$ for the ground 
state, and 
$\sigma_{\rm fus}(\theta)$ is the fusion cross section evaluated
with a potential given by Eq. (\ref{pot}) for a fixed value of $\theta$. 
Notice that $M$ is a projection of the angular momentum on the $z$-axis, 
which coincides with the beam direction at the initial stage of reaction. 
That is, the quantization axis for the ground state spin of the target nucleus 
is in the beam direction in this formula. 
Since the integrals of the $\phi$ and $\chi$ are trivial in Eq. 
(\ref{cross0}), 
one finally obtains, 
\begin{eqnarray}
\sigma_{\rm fus}^{(M)}(E)
&=&\frac{2I_0+1}{2}
\int^{\pi/2}_0\sin\theta d\theta \nonumber \\
&&\times\left(|d^{I_0}_{MK_0}(\theta)|^2
+|d^{I_0}_{M-K_0}(\theta)|^2\right)\, \sigma_{\rm fus}(\theta), 
\label{cross}
\end{eqnarray}
which coincides with Eq. (19) in Ref. \cite{Christley94} (see also
Ref. \cite{Hinde03}). 
Here we have used the relation 
$\sigma_{\rm fus}(\theta)=\sigma_{\rm fus}(\pi-\theta)$, 
which is valid for deformed nuclei with a reflection symmetric shape. 
Notice that, using the relation $\sum_M|d^I_{MK}(\theta)|^2=1$, 
fusion cross sections for the unpolarized target reads 
\begin{eqnarray}
\sigma_{\rm fus}^{(\rm unpol)}(E)
&=&\frac{1}{2I_0+1}\,\sum_M\sigma_{\rm fus}^{(M)}(E), \\
&=&
\int^{\pi/2}_0\sin\theta d\theta\,\sigma_{\rm fus}(\theta), 
\label{cross_unpol}
\end{eqnarray}
which is identical to the formula for even-even deformed nuclei \cite{HT12}. 

\begin{figure} [tb]
\includegraphics[scale=0.5,clip]{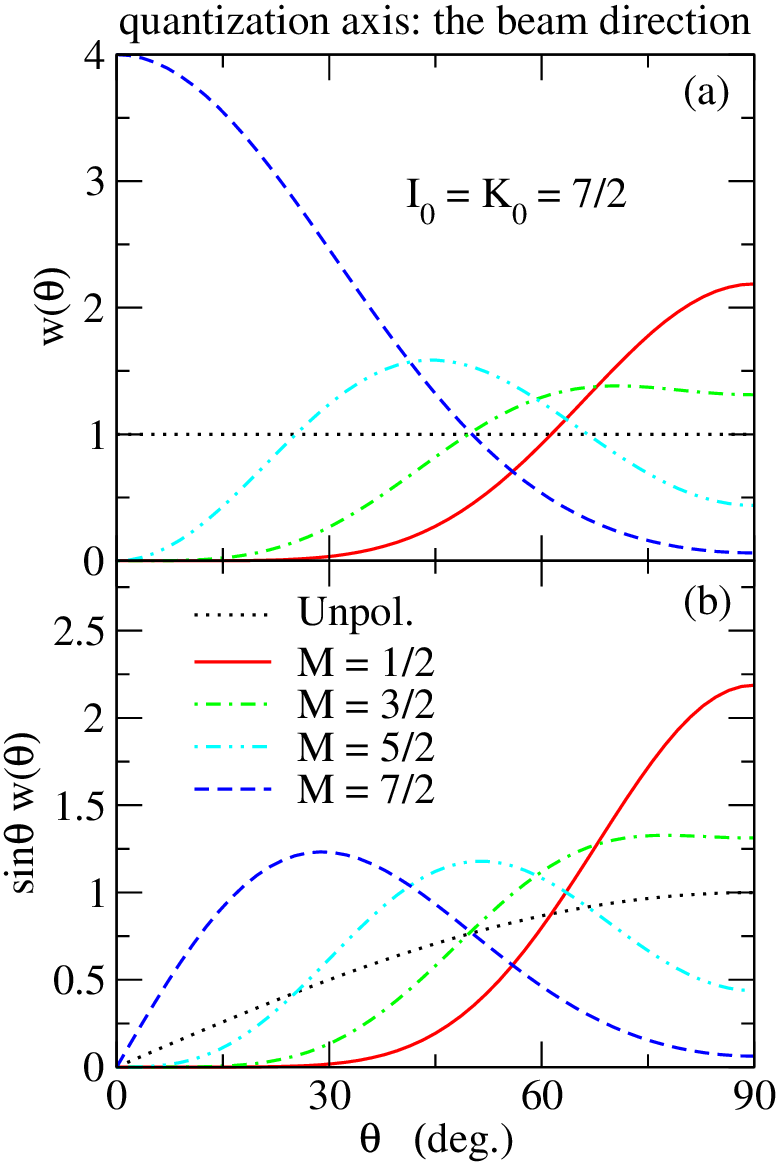}
\caption{
The weight factor for fusion cross sections of a deformed target 
nucleus with the ground state spin and parity of $I_0^\pi=7/2^-$ 
and the $K$ quantum number of $K_0=7/2$. 
It is plotted 
as a function of the angle between the symmetry axis of the 
target and the relative coordinate between the projectile and the target. 
The quantization axis is set to be parallel to the beam direction. 
The upper and the lower panels show the weight factor without and with 
the factor $\sin\theta$. 
The solid, the dot-dashed, the dot-dot-dashed, and the dashed lines 
are for $M=1/2,3/2,5/2$, and 7/2, respectively, where $M$ is the projection 
of the ground state spin of the target onto the quantization axis. 
The weight factor for the unpolarized case is also shown 
by the dotted 
lines. 
}
\end{figure}

Fusion cross sections for another direction of the quantization axis 
can also be computed by expanding the quantized state, $|I\tilde{M}\rangle$, 
with the eigenstates of $I_z$ as, 
\begin{equation}
|I\tilde{M}\rangle = \sum_Mc_M|IM\rangle. 
\end{equation}
Notice that the absolute value of the 
expansion coefficient, $c_M$, is actually given by 
\begin{equation}
|c_M|=|d^I_{\tilde{M}M}(\theta_a)|, 
\end{equation}
where $\theta_a$ is the angle between the quantization axis and the $z$-axis. 
Fusion cross sections are then given by, 
\begin{equation}
\sigma_{\rm fus}^{(\tilde{M})}(E)
=\sum_M|c_M|^2\sigma_{\rm fus}^{(M)}(E), 
\end{equation}
where $\sigma_{\rm fus}^{(M)}(E)$ is the fusion cross section 
when the quantization axis is taken to be the $z$-axis, 
given by Eq. (\ref{cross0}). 

\section{Application to the $^{16}$O+$^{165}$H\lowercase{o} system}

\begin{figure} [tb]
\includegraphics[scale=0.5,clip]{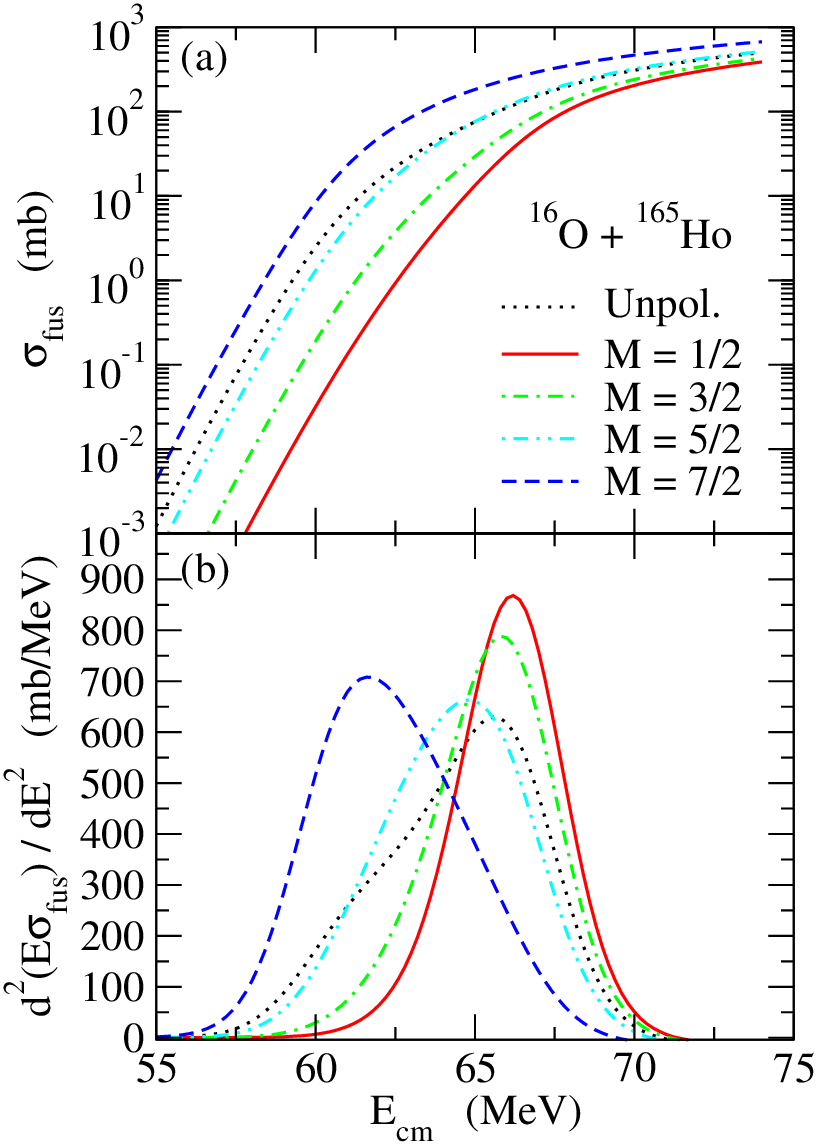}
\caption{
Fusion cross sections (the upper panel) and the fusion barrier distribution 
(the lower panel) for the $^{16}$O+$^{165}$Ho reaction. 
The meaning of each line is the same as in Fig. 1. 
}
\end{figure}

\begin{figure} [tb]
\includegraphics[scale=0.5,clip]{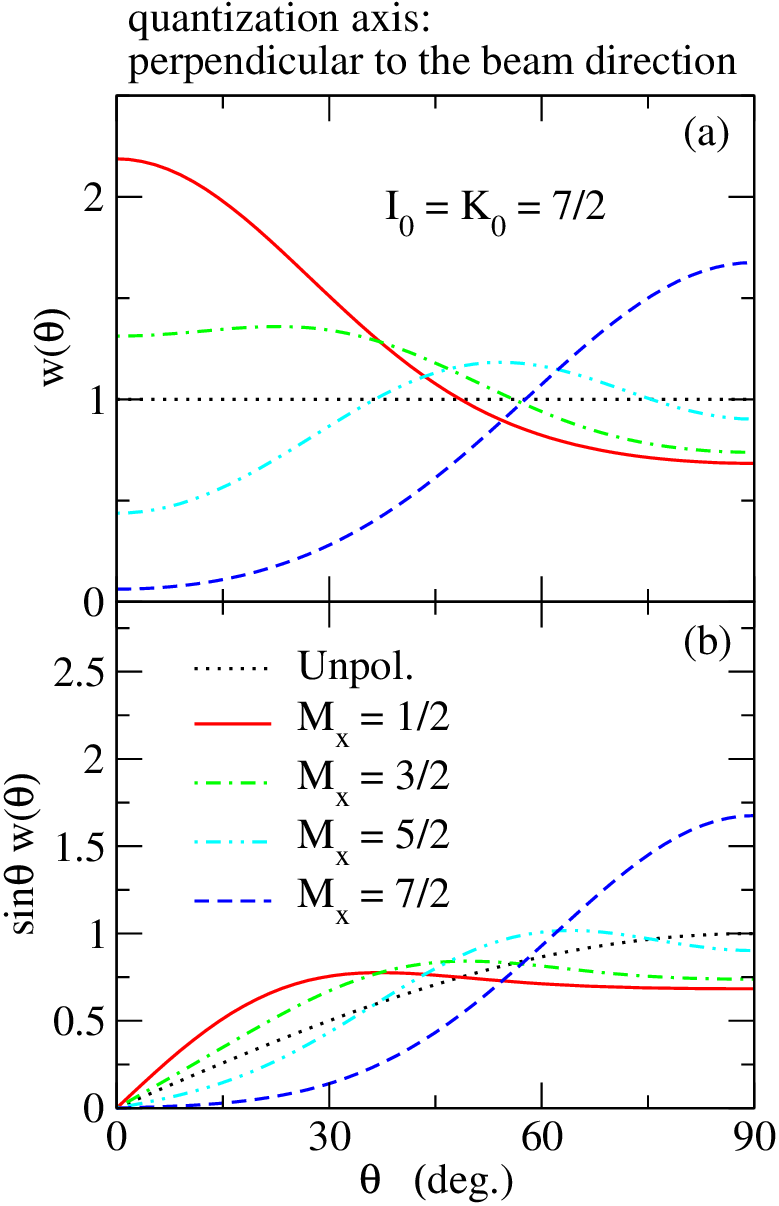}
\caption{
Same as Fig. 1, but for the case where the quantization axis is
perpendicular to the beam direction.}
\end{figure}

\begin{figure} [tb]
\includegraphics[scale=0.5,clip]{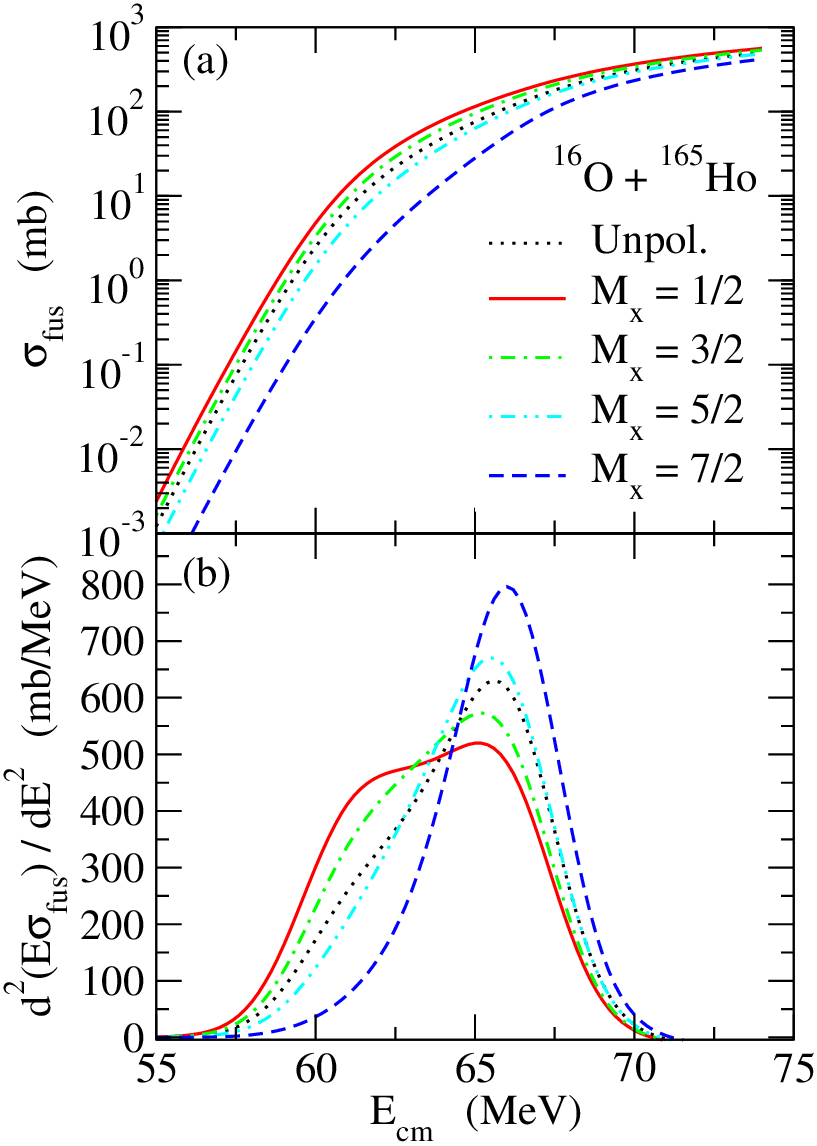}
\caption{
Same as Fig. 2, but for the case where the quantization axis is
perpendicular to the beam direction.}
\end{figure}

Let us now apply the formulas in the previous section to 
the $^{16}$O+$^{165}$Ho system and discuss the effect of alignment of the
target nucleus, $^{165}$Ho, that is a well deformed nucleus with
the ground state spin and parity of $I_0=7/2^-$.
The ground state rotational band has 
$K_0^\pi=7/2^-$. 
Notice that this nucleus has an ideal property as a material of the 
spin-aligned target. 
That is, due to its extremely strong hyperfine field, a large spin 
alignment of up to 80--90\% of the maximum theoretical value 
can be achieved by cooling down a single crystal of Ho metal 
\cite{Postma61,Wagner65,Kelly69,Stone68,Koster92,Huffman96}. 

We first take the quantization axis to be parallel to the beam direction.
Figure 1 shows the weight factor in the fusion cross sections
(see Eq. (\ref{cross})),
\begin{equation}
w_M(\theta)\equiv
\frac{2I_0+1}{2}
\left(|d^{I_0}_{MK_0}(\theta)|^2
+|d^{I_0}_{M-K_0}(\theta)|^2\right),
\label{weight}
\end{equation}
without (the upper panel) and with (the lower panel) the 
statistical factor of $\sin\theta$.
For comparison, the weight factor for the unpolarized case, $w(\theta)=1$, 
(see Eq. (\ref{cross_unpol})) is also shown.  
One can see that, whereas the distribution is isotropic for the unpolarized 
case, the side (the tip) collision is more emphasized for $M=1/2$ ($M=7/2$). 

In fusion reactions to synthesize superheavy elements, 
the side collision around $\theta\sim \pi/2$ 
contributes predominantly \cite{Hagino18,Tanaka18}. 
The figure implies that evaporation residue cross sections of 
superheavy nuclei could be increased by a factor of around 
two as compared to the unpolarized case, if the target nucleus 
could be selectively prepared with $M = \pm1/2$. 
In the synthesis of superheavy elements with extremely low 
cross sections as well as in fusion reaction with low intensity 
radioactive-ion beams, it is critically important to reduce beam 
times by making all possible efforts.
The enhancement by a factor of two suggested by this calculation
indicates that a spin alignment can be used for such 
purposes, even though a spin-aligned target which is applicable 
to fusion measurements will have to be developed.

Fusion cross sections for the $^{16}$O+$^{165}$Ho 
are shown in the upper panel of Fig. 2. 
To this end, 
we take the Woods-Saxon form for the internuclear potential, 
with the depth, the radius, and the diffuseness parameters of
$V_0=104$ MeV, $r_0=1.15$ fm, and $a=0.63$ fm, respectively. This potential
yields a similar barrier height as that with the Aky\"uz-Winther
potential \cite{AW81}. 
For the deformation parameters, we follow 
Ref. \cite{Moller16} and take 
$\beta_2=0.284$ and $\beta_4=0.020$, with the radius
parameter of $R_T=1.2A_T^{1/3}$ fm.
The excitation of $^{16}$O is taken into account only through the potential 
renormalization \cite{HT12,THAB94}, and is not explicitly included in the 
calculations. 
The figure corresponds well to Fig. 3 in Ref. \cite{Christley94}. 
Since the tip collision has a lower Coulomb barrier than the side 
collision, fusion cross sections for $M=7/2$ are much more enhanced 
as compared to those for $M=1/2$, reflecting the angle dependence of 
the weight factor as shown in Fig. 1. 

This fact can be seen more transparently in the fusion barrier distributions 
shown in the lower panel of Fig. 2. Here, the fusion barrier 
distribution, $D_{\rm fus}(E)$, 
is defined as the second energy derivative of 
$E\sigma_{\rm fus}(E)$ \cite{DHRS98,RSS91}, that is, 
\begin{equation}
D_{\rm fus}(E)=\frac{d^2(E\sigma_{\rm fus})}{dE^2}. 
\end{equation}
Here, we evaluate the barrier distributions using the point difference 
formula \cite{DHRS98} with the energy step of $\Delta E=2$ MeV. 
One can see that the shape of barrier distribution is considerably altered 
by the alignment, 
and moreover it is sensitive to the value of $M$, 
again by reflecting the angle dependence of the 
weight factor. 

Let us next discuss the case where the quantization axis of the ground 
state spin of the target nucleus is perpendicular to the beam direction, 
that is, the $x$ axis.  
Fig. 3 shows the weight factor, $\tilde{w}_{M_x}(\theta)$,  
\begin{equation}
\tilde{w}_{M_x}(\theta)=\sum_M|d^{I_0}_{M_xM}(-\pi/2)|^2\,w_M(\theta), 
\label{weight_x}
\end{equation}
where $w(\theta)$ is the weight factor given by Eq. (\ref{weight}). 
Here we have used the notation $M_x$ to denote
the magnetic substates in this case, in order to 
distinguish them from those in the case where the 
quantization axis is along the beam axis. 
One can see that the role of $M=1/2$ and $M=7/2$ is inverted from the 
case where the quantization axis is parallel to the beam axis (see Fig. 1). 
That is, the tip (the side) collision is more emphasized 
for $M_x=1/2$ ($M_x=7/2$), that is opposite to Fig. 1. 
For $M_x=7/2$, even though the weight at $\theta=\pi/2$ is somewhat 
reduced compared to the weight factor for $M=1/2$ shown in Fig. 1, 
it is still significantly 
larger than the weight factor for the unpolarized case, by a factor 
of about 1.5. This would be an important implication for evaporation 
residue cross sections for superheavy nuclei. 

\begin{figure} [tb]
\includegraphics[scale=0.5,clip]{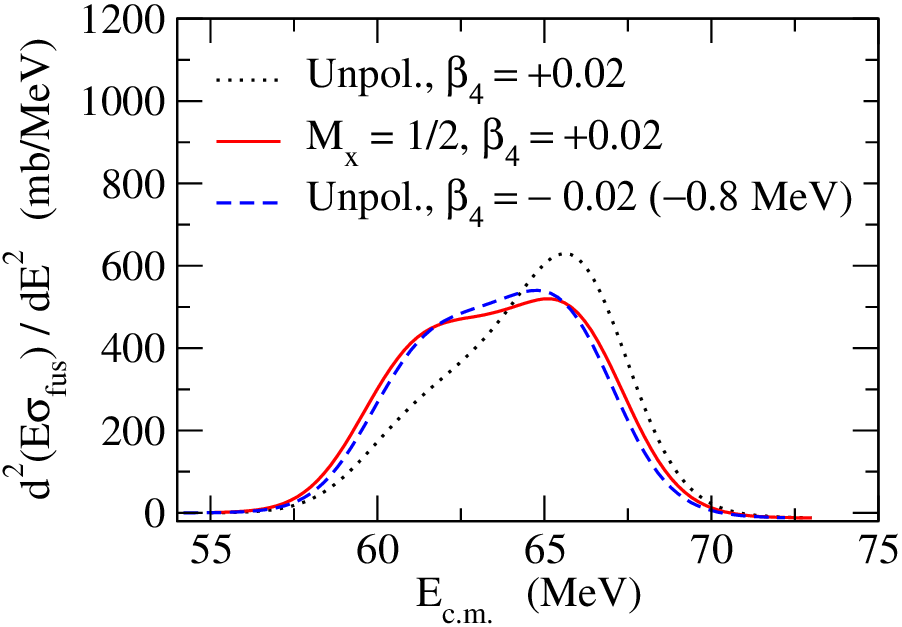}
\caption{Comparison of the fusion barrier distributions for the 
$^{16}$O+$^{165}$Ho reaction obtained with several schemes. 
The meaning of the dotted and the solid lines is the same as 
in Fig. 4 (b), that is, the dotted line denotes the barrier distribution 
for the unpolarized case while the solid line shows the result 
for $M_x=1/2$ with the quantization axis being perpendicular to the 
beam direction. The dashed line shows the barrier distribution for the 
unpolarized case, but by changing the sign of the hexadecapole deformation 
parameter of $^{165}$Ho. To facilitate the comparison, the dashed line is 
shifted in energy by $-0.8$ MeV. 
}
\end{figure}

The fusion cross sections and the fusion barrier distributions are shown 
in Fig. 4. The fusion cross sections for $M_x=7/2$ are 
considerably smaller than those for the other values of $M_x$, 
because this configuration contains much smaller component of the tip 
collision ($\theta\sim 0$), 
which has the lowest Coulomb barrier, compared to the other 
configurations (see Fig. 3). As in the case where the quantization axis 
is parallel to the beam axis, one can see that 
the shape of barrier distribution depends sensitively on the 
value of $M_x$. Interestingly, the shape of barrier distribution 
for $M_x=1/2$ is a typical one for an even-even deformed nucleus 
with $\beta_2>0$ and $\beta_4<0$ \cite{DHRS98,LDH95}. 
To demonstrate this, 
Fig. 5 compares the barrier distribution for $M_x=1/2$ (the solid line) 
with that for the 
unpolarized case obtained by inverting the sign of $\beta_4$ (the dashed 
line). To facilitate the discussion, 
the solid line is shifted in energy by $-0.8$ MeV. 
For a comparison, the figure also shows the barrier distribution for the 
unpolarized case with the positive sign of $\beta_4$ (the dotted line). 
In addition to the well known fact that the shape of barrier distribution 
is sensitive to the sign of $\beta_4$ \cite{DHRS98,LDH95}, one can see 
that the solid line is indeed similar to the dashed line. 
This may suggest that the hexadecapole deformation can be effectively changed 
rather arbitrarily in heavy-ion fusion reactions 
by appropriately aligning an odd-mass target, even though the intrinsic hexadecapole 
deformation itself remains the same. 

\section{Summary}

We have discussed the role of alignment in heavy-ion subbarrier fusion 
reactions of a well-deformed odd-mass nucleus. Such nucleus has a finite 
spin in the ground state, 
and fusion cross sections differ depending on how the nucleus is polarized. 
We have in particular considered the $^{16}$O+$^{165}$Ho system, 
and employed the iso-centrifugal and the sudden tunneling approximations. 
With these approximations, the fusion cross sections are given 
as a weighted average of 
orientation dependent cross sections, in which the weight factor is 
given in terms of the ground state wave function of the target nucleus 
for each quantum number $M$, $M$ being the projection of the 
initial spin of the deformed target nucleus on to the direction of the 
beam axis. We have shown that the fusion cross sections and the shape of 
barrier distribution is sensitive to the magnetic substate of the target 
nucleus. In particular, whereas $\beta_4$ is positive in $^{165}$Ho, 
the shape of barrier distribution becomes similar 
to a typical one for an even-even deformed nucleus with 
a negative value of $\beta_4$, 
when the odd-mass target is aligned along the 
axis perpendicular to the beam axis. This may imply that 
one can control the hexadecapole deformation in subbarrier fusion 
reactions by aligning deformed target nuclei. 
We have also pointed out that fusion 
cross sections for the side collision can 
be enhanced by a factor of around 2, by selectively choosing a particular 
value of a magnetic substate. This would be a good advantage for 
synthesizing superheavy elements, for which the side collision predominantly 
contributes. Moreover, use of a spin-aligned target would enable one to control the 
orientation of the deformed nuclei and to investigate the dynamics 
of fusion reactions by decomposing tip and side contributions.

\section*{Acknowledgments}
We thank Phil Walker for useful conversations. This work was supported by
JSPS KAKENHI Grant Number JP19K03861.

\end{document}